\documentclass[preprint,10pt]{JHEP3} 

\JHEPspecialurl{http://jhep.sissa.it/JOURNAL/JHEP3.tar.gz}

\usepackage{epsfig,multicol,amsmath}

\usepackage{epstopdf}

\def\beq{\begin{equation}}
\def\eeq{\end{equation}}
\def\bea{\begin{eqnarray}}
\def\eea{\end{eqnarray}}

\def\eq#1{{Eq.~(\ref{#1})}}
\def\fig#1{{Fig.~\ref{#1}}}
\newcommand{\bas}{\bar{\alpha}_S}
\newcommand{\as}{\alpha_S}

\newcommand{\Lb}{\left(}
\newcommand{\Rb}{\right)}

\parskip=\itemsep               

\def\pom{{I\!\!P}}

\newcommand{\h}{\frac{1}{2}}

\newcommand{\cpom}{ \displaystyle{\not }\pom}

\def\thefootnote{\fnsymbol{footnote}}
\def\blfootnote{\xdef\@thefnmark{}\@footnotetext}

\title{\huge \bf Multiparticle production in the mean field approximation of high density QCD}
\author{\Large  Andrey~Kormilitzin\thanks{ andreyk1@post.tau.ac.il}, Eugene~Levin   \thanks{ leving@post.tau.ac.il,
levin@mail.desy.de;} \,\,and\,\,Alex~Prygarin
 \hspace{0.001cm}
\thanks{ prygarin@post.tau.ac.il
}\\
Department of Particle Physics, School of Physics and Astronomy\\
Raymond and Beverly Sackler
 Faculty
of Exact Science\\  Tel Aviv University, Tel Aviv, 69978, Israel
}

\abstract{The generating functional is suggested for multiparticle generation processes.  In mean field approximation  of high density QCD two equations for new generating functional are derived: linear functional equation for an arbitrary initial condition and non-linear  one for a specific initial condition. The non-linear equation has the form of Kovchegov-Levin equation for diffraction production and gives its generalization on
the processes with fixed multiplicities of produced particles.  }


\preprint{  TAUP-2879/08\\
 \today}

\begin{document}

\def\thefootnote{\arabic{footnote}}
\section{Introduction:\,   AGK cutting rules}\label{sec:Int}
The goal of this paper is to develop a technique that will allow us to calculate more exclusive processes than the total cross section in the framework of high density QCD \cite{GLR,MUQI,MV}. We will work in  the  mean field approximation in which we have Balitsky-Kovchegov\cite{B,K} non-linear evolution equation   for the elastic high energy amplitude.  Except the obvious motivation for expanding the calculative power of high density QCD we have a more specific   reason for doing this paper: to give more transparent derivation of the 
Kovchegov-Levin equation \cite{KLDD} for the diffractive dissociation cross section in the mean field approximation\footnote{One of us (E.L.), has heard a lot of complaints from experts that Ref.\cite{KLDD} is impossible to understand.}. 

The tool which we are going to use, is the AGK gutting rules \cite{AGK}.  In high energy phenomenology  these rules are very useful in calculation of the processes of different multiplicity from the expression for the total cross section.   Since we have a very powerful framework for high density phenomena in QCD, namely,  the dipole approach \cite{MUCD},  we wish to expand this approach to multiparticle production.   The AGK cutting rules, being proved in QCD, will  allow us to approach such  exclusive processes  as diffractive production and different correlations in multiparticle production processes. 
For a long time the situation with the proof of the AGK cutting rules has been uncertain (see Refs.\cite{GLR,AGKQCD,K1,K2,K3}). At first sight in the leading $\log(1/x_{Bjorken})$ approximation \cite{BFKL} of perturbative QCD the scattering amplitude, as a function of particle masses, decreases     enough to apply  
the original arguments of the AGK paper \cite{AGK},  on the other hand, the main question whether the total cross section and the multiparticle production can be described by the same set of diagrams, remains unanswered.  The situation became even worse when the explicit violation of the AGK cutting rules were found in  Refs. \cite{K1,K2,K3}. Fortunately, we believe that the mess with AGK cutting rules in QCD has been resolved in Ref. \cite{LEPR}.  However,  before describing the main results of this paper which we will use below, we give an introduction to AGK cutting rules explaining what they claim.

In simple language,  the AGK cutting rules give us the relation between the total cross section at high energy and the processes of multiparticle production. In QCD the main idea stems from the unitarity constraint for the BFKL Pomeron \cite{BFKL}, which describes the high energy scattering amplitude in the Leading Log $(1/x)$ Approximation of perturbative QCD. The unitarity  reads as
\beq \label{UNC}
2 N\Lb Y; x,y\Rb \,\,\,=\,\,| N\Lb Y; x,y\Rb|^2 + G_{in}\Lb Y; x,y\Rb
\eeq
where $Y=\log (1/x_{Bjorken})$
 and $(x,y)$ are the coordinates of the incoming dipole; $ N\Lb Y; x,y\Rb$ is the imaginary part of the  elastic amplitude and the first term describes the elastic scattering (assuming that the real part of the amplitude is small at high energy), while the second term stands for the contribution of all inelastic processes. In the leading log $(1/x)$ approximation the elastic contribution can be neglected and for the BFKL Pomeron \eq{UNC} can be reduced to the form (see \fig{cutpom}):
\beq \label{UNPO}
2 N^{BFKL}\Lb Y; x,y\Rb \,\,=\,\, G^{BFKL}_{in}\Lb Y; x,y\Rb
\eeq
We call $G_{in}\Lb Y; x,y\Rb$ by cut Pomeron while $N^{BFKL}\Lb Y; x,y\Rb$ will be called Pomeron or uncut Pomeron.
\FIGURE[ht]{\begin{minipage}{85mm}
\centerline{\epsfig{file=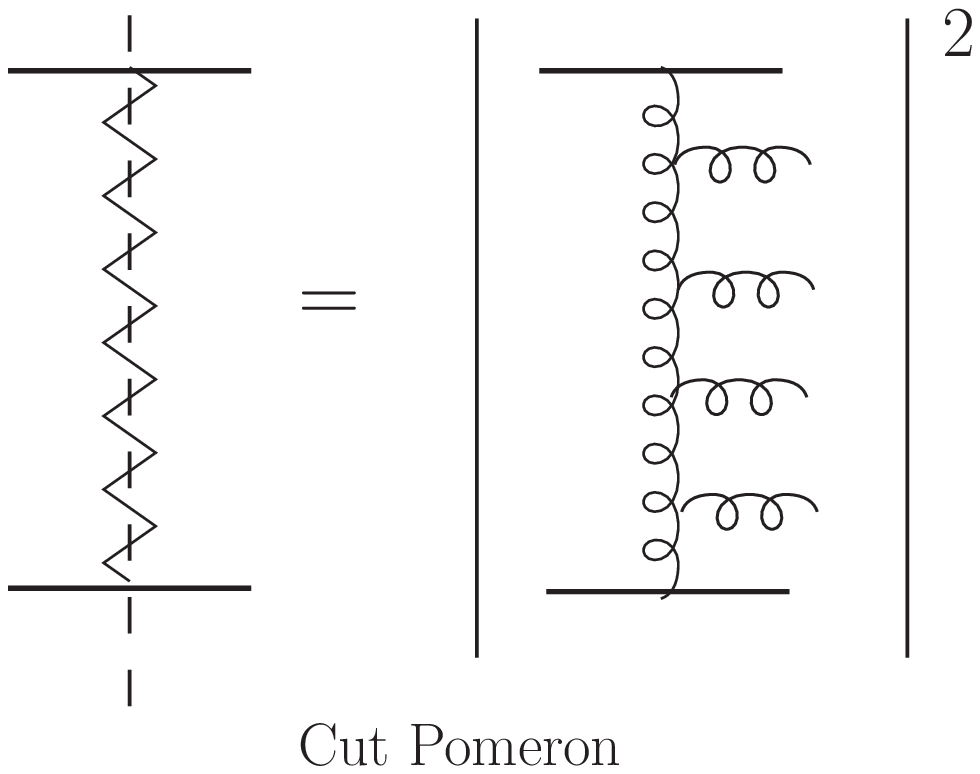,width=80mm,height=40mm}}
\end{minipage}
\caption{The definition   of cut Pomeron through the BFKL ladder. }
\label{cutpom} }

The original AGK cutting rules state that if we  know  the contribution to the total cross section of  the exchange of any number of Pomerons we  can calculate the processes with different multiplicity. In \fig{agkin} you can see the simple triple Pomeron diagram with the AGK coefficients (see \fig{agkin}-a). The coefficients mean that you need to multiply by these coefficients the contribution of triple Pomeron diagram in the total cross section to obtain the cross section of the multiparticle production.  In Ref.~\cite{LEPR} it was proved that these AGK cutting rules are correct in QCD for the triple BFKL Pomeron vertex.

\FIGURE[h]{ \begin{tabular}{c}
\centerline{\epsfig{file= 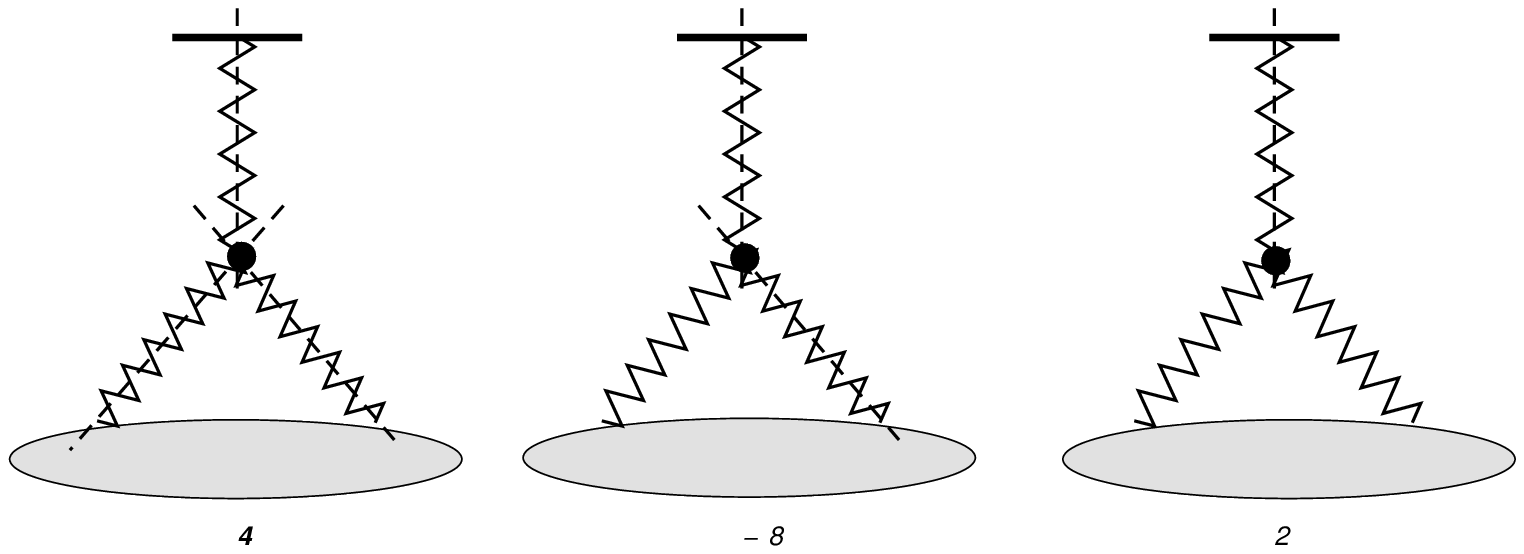,width=140mm,height=30mm }} \\
\fig{agkin}-a\\
\centerline{\epsfig{file= 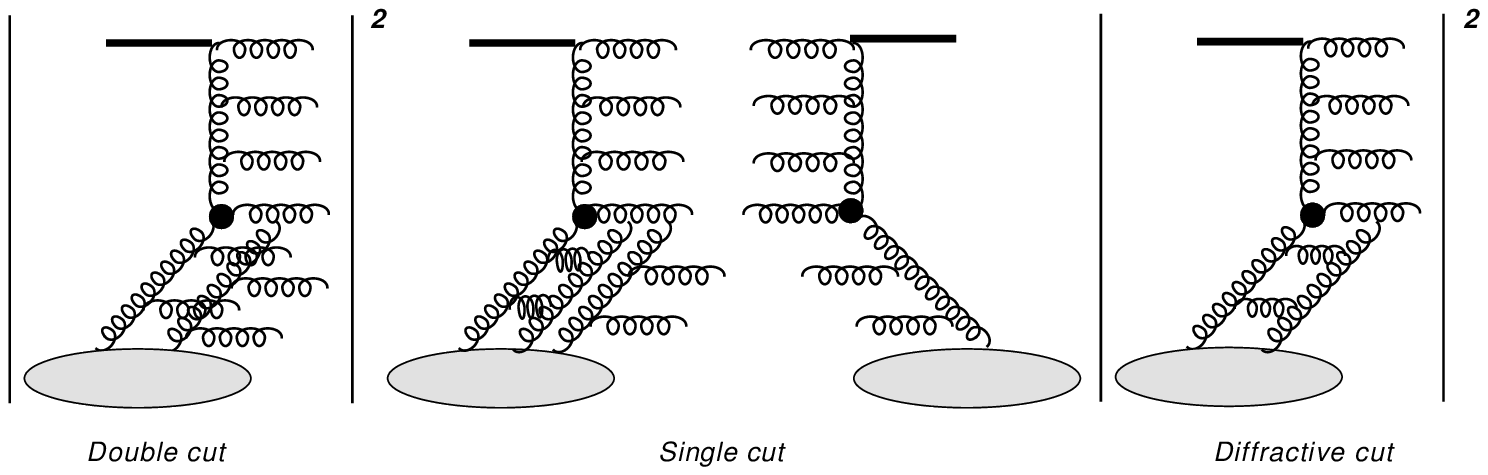,width=140mm,height=30mm }} \\
\fig{agkin}-b\\
\end{tabular}
\caption{ The AGK cutting rules for  the total inclusive processes. Cut Pomeron is defined in \protect\fig{cutpom} and in \protect\eq{UNPO}. In \fig{agkin} one can see decoding of the Pomeron diagrams in terms of the production processes.
} \label{agkin} }
 The exact form of the AGK rules look as follows.
\begin{eqnarray}
\mbox{Double Cut:} & \,\,\,\,\,\,\,\,\,\,\,\,&\frac{\bas}{2 \pi} \int\,d^2 x_2\,K\Lb x_{10}|x_{02},x_{12} \Rb
\left\{2\,\Lb N_{BFKL}\Lb 12\Rb\,\,+\,\,   N_{BFKL}\Lb 02\Rb \Rb^2\,\,-\,\,2\,  N^2_{BFKL}\Lb 10\Rb \right\}
\label{DC}
\\
\mbox{Single Cut:} &\,\,\,\,\,\,\,\,\,\,\,\,&
 \frac{\bas}{2 \pi} \int\,d^2 x_2\,K\Lb x_{10}|x_{02},x_{12} \Rb \left\{
-\,4\,\Lb N_{BFKL}\Lb 12\Rb\,\,+\,\,   N_{BFKL}\Lb 02\Rb \Rb^2\,\,+\,\,4\,  N^2_{BFKL}\Lb 10\Rb
 \right\} \label{SC}
\\
\mbox{Diffractive Cut:} &  &
 \frac{\bas}{2 \pi} \int\,d^2 x_2\,K\Lb x_{10}|x_{02},x_{12} \Rb \left\{
\Lb N_{BFKL}\Lb 12\Rb\,\,+\,\,   N_{BFKL}\Lb 02\Rb \Rb^2\,\,-\,\, N^2_{BFKL}\Lb 10\Rb \right\}\label{DDC}
\end{eqnarray}
where the vertex of decay of one dipole $x_{10}$ to two dipoles with sizes $x_{02}$ and $x_{12}$
is equal to
\beq \label{K}
\frac{\bas}{2 \pi}\,\,K\Lb x_{10}|x_{02},x_{12} \Rb\,\,=\,\,\frac{\bas}{2 \pi}\,\frac{x^2_{10}}{x^2_{02}\,x^2_{12}}
\eeq
$N_{BFKL}\Lb x_{ik}\Rb$ is the scattering amplitude of the dipole with size $x_{ik}$ off the target in the leading $\log(1/x_{Bjorken})$ approximation of perturbative QCD (the BFKL Pomeron exchange).
It is clear from the above equations that they have a simple meaning: the dipole with size $x_{10}$ decays in two dipoles with sizes $x_{02}$ and $x_{12}$ in the case when  both of them  interact with the target with multiparticle production ( double cut of \eq{DC}); one interacts with the target with multiparticle  production while the second
one  interacts elastically (single cut of \eq{SC}); and  two produced dipoles scatter elastically (diffractive cut of \eq{DDC}).

It turns out that difficulties with the AGK cutting rules that has been discussed,  stem
from the fact that in QCD the AGK cutting rules do not work for the situation when we an extra gluon is emitted from the vertices (see \fig{agkver}).
\FIGURE[h]{ \centerline{\epsfig{file= 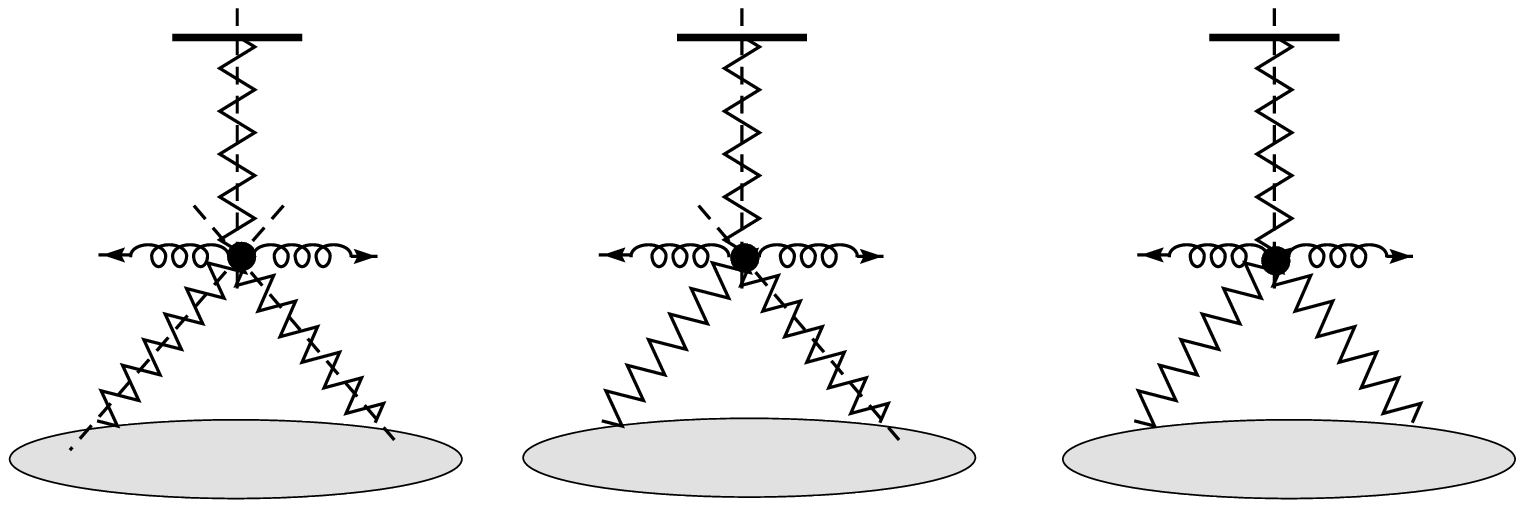,width=120mm,height=25mm}}
\caption{ Violation of AGK cutting rules for the processes with additional emission of one particle (gluon). Cut Pomeron is defined in \protect\fig{cutpom} and in \protect\eq{UNPO}.
} \label{agkver} }
The difference between these two cases are due to the fact that  for multiparticle production without measuring the gluon, is described by the same set of the diagrams as the total cross section, while for the events with measured gluon the set of diagrams turns out to be quite different (see Ref. \cite{LEPR}).

In the next section we will derive the generating functional for multiparticle production in  the dipole approach.
Therefore, for the dipole approach to QCD we will repeat the program that has been worked out for the 
BFKL Pomeron Calculus in zero transverse dimension \cite{LEPR0}.  In section 3 we derive the non-linear equation for the generating functional and for the cross sections of multiparticle production.
These equations will lead to a natural generalization of the Kovchegov-Levin equation for the diffractive production cross section (see Ref. \cite{KLDD}). In conclusions we summarize our results.

\section{Generating functional for multiparticle production processes}\label{sec:genfun}
\subsection{Generating functional for total cross section}\label{sec:genfuntot}
In the mean field approximation (MFA) we take into account only one Pomeron to two Pomeron splitting neglecting  merging of two Pomerons into one Pomeron. In other words,  we consider only `fan' diagrams.
The simple process for which such approximation can work, is the deep inelastic scattering with a nuclear target (see Ref.\cite{K}).  In the dipole approach,  one Pomeron to two Pomeron splitting reduces to the decay
of one dipole to two dipoles with the probability of this decay given by \eq{K}\cite{MUCD}.
The simplest and the most transparent technique to incorporate this decay is the generating functional which allows us to reduce the calculation of the high energy elastic amplitude to consideration of a Markov process, The advantage of this approach is the fact that it takes into account both $t$ and $s$-channel unitarity as we will discuss below.

The  generating functional is defined as\cite{MUCD,LL}
\beq \label{Z11}
Z_0\left(Y -y;\{u\} \right)\,\,\equiv\,\,\sum_{n=1}\,\int\,\,
P_n\left(Y -y;r_1, \dots  ,r_n \right) \,\,\prod^{n}_{i=1}\,u(r_i) 
d^2r_i
\eeq 
where $u(r_i)$ is an arbitrary function of $r_i$ and $b_i$. $P_n$ is the probability density to 
find  $n$ dipoles with sizes $r_1,\dots,r_n$ at  rapidity $Y-y$. 
 \footnote{The dipole $(x_i,y_i)$ with coordinates $x_i$ for quark  and $y_i$ for antiquark can be characterized by the dipole size $\vec{r}_i = \vec{x}_i  - \vec{y}_i$ and $ \vec{b}_i = \h ( \vec{x}_i  + \vec{y}_i)$ . For simplicity we suppress in \eq{Z11} and below the coordinate $b_i$. For the scattering with the nuclear target we can consider  that impact parameters of all dipoles are the same $b_i = b$ (see Ref. \cite{K}).}
For functional of \eq{Z11} we have two obvious  conditions:
\begin{itemize}
\item \quad at $Y =y$ \,\, $P_n\,=\,0$ for $n \,>\,1$ and $P_1 = \delta( \vec{r} 
\,-\,\vec{r}_1)$. In other words, at $y =Y$ we have one dipole of size $r$
\beq \label{ZIC}
Z_0\left(Y -y=0;\{u\}\right)\,\,=\,\,u(r)\,\,;
\eeq
\item \quad at $u =1$ 
\beq \label{ZBC} 
Z_0\left(Y -y;\{u\}\right)|_{u=1}\,\,=\,\,1.
\eeq
\eq{ZBC} follows from the physical meaning of $P_n$ and represents the conservation of the total probability.
\end{itemize}

For probabilities $P_n$ we can write the following equation
\bea \label{Z2}
&&- \,\,\frac{d P_n\left(Y -y;r_1,\dots, r_i, \dots , r_n \right)}{d 
y}\,\,\,= \\
&&\,\,-\,\,
\,\frac{\bas}{2 \pi}\,\sum^n_{j=1}\;\int\,d^2 r' 
\,K\Lb r_j; r', |\vec{r}_j - \vec{r'}|\Rb
P_n\left(Y -y;r_1,\dots ,r_j, \dots,  r_n 
\right)\,\,\notag \\
&&+\,\,\frac{\bas}{2 \pi}\, \,\,\sum^{n-1}_{j=1}
\;\int\,d^2 r' d^2 \tilde{r}_j 
\delta(\vec{\tilde{r}}_j-\vec{r}_i+\vec{r}_n)
\delta(\vec{r'}-\vec{r}_n)
\, \,K\Lb \tilde{r}_j; r', |\vec{\tilde{r}}_j - \vec{r'}|\Rb  
\,P_{n - 1}\left(Y -y;r_1, \dots, \tilde{r}_j, \dots , r_{n-1}
 \right)\notag
\eea
\eq{Z2} describes a typical Markov process:  two terms of this equation has simple meaning of 
increase of  the probability to find $n$-dipoles due to decay of one  dipole to two dipoles (birth terms, the second term in \eq{Z2}) and of decrease of the probability since one of $n$-dipoles can decay (death term, the first tern in \eq{Z2}).
Multiplying by product $\prod^n_{i=1} u(r_i)$ and integrating over $r_i$ we obtain the 
following linear equation for the generating functional
\bea \label{Z3}
\frac{d Z_0\left(Y -y;\{u\} \right)}{d
y}= 
 & & \,\frac{\bas}{2 \pi}\,\int\,\,d^2 r \;d^2 r' \,K\Lb r; r', |\vec{r} - \vec{r'}|\Rb \left\{-\,u(r)\,
+\,
u(r')\,u(|\vec{r'} - \vec{r}|)\right\}
\,\frac{\delta}{\delta u(r)} \,\, Z_0\left(Y -y;\{ u\} \right)
\eea

Here we use notation $\delta/\delta u(r)$ for the functional derivative.
Using the generating functional we can calculate the scattering amplitude $N(Y,r,b)$ as follows \cite{LL}

\bea
N(Y-Y_0,r,b) \,\,&=&\,\,- \sum^{\infty}_{n=1}\,\,\frac{ 
(-1)^{n+1}}{n!}\,\,\left\{\int\,\prod^n_{i =1}\,\,d^2r_i\,\,\gamma(r_i)\, \frac{\delta}{\delta 
u(r_i)} \,\,\right\} \,Z_0\left(Y -Y_0;\{u\} \right)|_{u = 1} \label{N1}\\
 &\equiv & \,1 \,\,-\,\,Z_0\left(Y -Y_0;\{u\} \right)|^{u(r)}_{u=1-\gamma} \label{N11}
\eea
where $ \gamma(r) \,=\,N(Y_0,r,b) $ is the dipole scattering amplitude at low energy ($Y_0$).
The notation $Z_0\left(Y-Y_0;\{u\} \right)|^{u(r)}$ means that  the initial condition of the generating functional is given by \eq{ZIC}. In a more general case we can define the scattering amplitude for any arbitrary initial condition $F(u(r))$, for example, if we choose $F(u(r))=u^2(r)$, we start with two dipoles at the initial rapidity $Y$.
  In \eq{N1} we assume that the low energy amplitude of interaction of $n$ dipoles with the target  $\gamma_n$ is equal to $\gamma_n = \gamma^n$, which means that dipoles interact with the target independently (without correlations). 

\eq{Z3} being a linear equation with only first derivative has a general solution of the form 
$Z_0\left( Y-y;\{u\}\right) \equiv 
Z_0\left(\{u(Y-y)\} \right)$. Inserting this solution and using the initial condition of \eq{ZIC} we obtain the non-linear equation
\small
\beq  \label{Z4}
\small
\frac{d Z_0\left(Y-y;\{u\}\right)|^{u(r)}}{dy}=\frac{\bas}{2 \pi}\int d^2 r'
 \,K\Lb r;r',| \vec{r} - \vec{r'}|\Rb \left\{-Z_0\left(Y-y,\{u\}\right)|^{u(r)}+
Z_0\left(Y-y,\{u\}\right)|^{u(r')}\,Z_0\left(Y-y,\{u\}\right)|^{u(|\vec{r}-\vec{r'}|)}\right\}
\eeq
\normalsize
It is easy to see that using \eq{N11} for the amplitude \eq{Z4} can be re-written as the Balitsky-Kovchegov  equation \cite{B,K} in the form
\bea \label{N2}
\frac{d N\left(Y-y,r,b\right)}{d(Y-y)}\,\,&=&\,\,\frac{\bas}{2 \pi}\,\int\,\,d^2 r' \,K\Lb r;r',| \vec{r} - \vec{r'}|\Rb \times \\ 
& \times &\left\{
N\left(Y-y,r',b\right) \,+\,N (Y-y,|\vec{r} - \vec{r'}|,b )\,-\,N\left(Y-y,r,b\right) \,-\,N\left(Y-y,r',b\right)\,N (Y-y,|\vec{r} - \vec{r'}|,b )\right\}\notag
\eea

The evolution equation in its linear form \eq{Z3}    has two advantages in comparison with
the non-linear equations   \eq{N1}: it has a simple statistical interpretation and can be solved with arbitrary initial condition while non-linear equation are correct only with specific initial condition of \eq{ZIC}.
 The statistical interpretation allows us to reduce the problem of $s$-channel unitarity to the conservation of probabilities at each level of rapidity $y$ which is included in  Markov chain equation (see \eq{Z2} ). Since the technique is equivalent to summing of the Pomeron diagrams the $t$-channel unitarity is preserved, but we need to remember that we must sum all Pomeron diagrams to fulfill $t$-channel unitarity including the so called Pomeron loops. 

The equivalence between the generating functional approach and  the Pomeron Calculus becomes clear if we introduce a new functional
\beq \label{Nfundef}
N\Lb  Y-y,\{\gamma\}\Rb \,\,=\,\,1 - \,\,Z_0\left(Y-y;\{u\}\right)|_{u=1-\gamma}
\eeq
 for which we have the equation
\bea \label{N3}
\frac{d\, N\left(Y-y;\{\gamma\} \right)}{d
(Y-y)}= \frac{\bas}{2 \pi}\int d^2 r \;d^2 r' K\Lb r;r',| \vec{r} - \vec{r'}|\Rb \left\{
 \gamma(r')+\gamma(\vec{r} - \vec{r'})-\gamma(r)\,
-
\gamma(r')\,\gamma(\vec{r} - \vec{r'})\right\}
 \frac{\delta}{\delta \gamma(r)}   N\left(Y-y,\{\gamma \} \right) \hspace{0.3cm} 
\nonumber \\
\eea
Starting with the initial condition 
\begin{eqnarray}\label{initN}
N\left(Y_0;\{\gamma \} \right) = \gamma(r)
\end{eqnarray} 
found from \eq{ZIC} and \eq{N11}, one can easily built the Pomeron Calculus iterating \eq{N3}.

\subsection{Generating functional for multiparticle production: general ideas}
In the spirit of Ref.~\cite{LEPR0} we wish to introduce a generalization of \eq{Z11}, namely,
\bea \label{Zmulti}
 Z \left(Y-y;\{u\} ,\{v\}\right)\,\,\equiv 
  \,\,\sum^{\infty}_{n=1,m=0}\,\int\,\,
P^m_n\left(y;r_1, \dots ,r_n; r_1,\dots , r_m \right) \,\,\prod^{n}_{i=1}\,u(r_i) \prod^{m}_{k=1}\,v(r_k) 
d^2r_i\,d^2r_k 
\eea
where $P^m_n$ is the probability to find $n$ uncut Pomerons and $m$ cut Pomerons. 
In our discussion we use the fact that colorless dipoles are proper degrees of freedom for partonic description 
of the BFKL Pomeron.
To treat this probability in terms of the dipole approach we need to introduce two typical moments of time for the processes of multiparticle production:  $\tau =0$ when the interaction with the target occurs, and $\tau = \infty$ where our detectors for particles are located.  Having in mind these two moments of time  
we can 
treat $P^m_n$ as the probability to have at rapidity $y$ $n$ dipoles with sizes $r_1, \dots, r_n$ at $\tau = 0$, which do not survive until $\tau=\infty$ and cannot be measured, while the dipoles with sizes $
r_1, \dots, r_m$  reach $\tau = \infty$ and can be caught by detectors. 

The first boundary condition for the generating functional is obvious
\beq \label{MPZBC}
Z \left(Y-y;\{u\} ,\{v\}\right)|_{u=1,v=1} \,\,=\,\,1
\eeq
and it follows from the physical meaning of $P^m_n$ and represents the conservation of the total probability at any rapidity. 

Generalizing \eq{N1} we can introduce unintegrated over impact parameter cross section for 
multiparticle production as
\bea
&&M\Lb Y-Y_0;r,b\Rb \,\,\equiv \label{MP}\\
&&
\sum^{\infty}_{n=1,m=0}\,\,\frac{ 
(-1)^{n + m + 1}}{n!\,m!}\,\,\left\{\int\,\prod^n_{i =1}\,\,d^2r_i\,\int\,\prod^m_{k =1}\,\,d^2r_k\,\gamma(r_i)\,\gamma_{in}(r_k) \,\,
\frac{\delta}{\delta u(r_i) } \,\,\frac{\delta}{\delta v(r_k) }\right\} \,Z\left(Y-Y_0); \{u\} \right)|^{v(r)}_{u = 1,v=1} \notag\\
&&  \,\,\,\,\,\,\,\,\,\,\,\,\,\,\,\, =   1\,\,\,-\,\, Z\Lb Y-Y_0; \{u\},\{ v\}  \Rb|^{v(r)}_{u=1-\gamma,v=1-\gamma_{in}}\notag
\eea
The total cross section is obtained from  \eq{MP}  substituting $2 \gamma( r ) \,=\,\gamma_{in}(r )$, where $\gamma( r)$ is the low energy ($Y_0$) amplitude of the  dipole interaction with the target.   $2 \gamma( r) \,=\,\gamma_{in}( r)$ follows directly from the unitarity constraint of \eq{UNPO} at low energies. We want to remind at this point that  we  consider $Y_0$ is so large that 
$\as Y_0 \approx 1$ and we can use the Born approximation of perturbative QCD for estimates of $\gamma_{in}(r)$.  Using new notation for elastic amplitude $N$ (see \eq{N1}) and generating functionals $Z_0$ and $Z$  (see \eq{Z11} and \eq{Zmulti}),  we can re-write \eq{UNC} in the form
\beq \label{MP111}
2 N(Y-y,r,b)=  2\,\Lb 1 - Z_0\Lb Y-y; \{u \}\Rb|^{u(r)}_{u=1-\gamma}\Rb\,\,=\,\,|N(Y-y,r,b)|^2 + G_{in} \,=\,\,
   1\,\,\,-\,\, Z\Lb Y-y; \{u\},\{ v \}\Rb |^{v(r)}_{u=1-\gamma,v=1-2\gamma}
\eeq
which translates into boundary condition 

\beq \label{MP1}
Z\Lb Y-y; \{u\},\{ v \}\Rb |_{v=2u-1} =2 \;Z_0\Lb Y-y; \{u \}\Rb\,\,-1
\eeq

\FIGURE[h]{\begin{minipage}{70mm}
\centerline{\epsfig{file=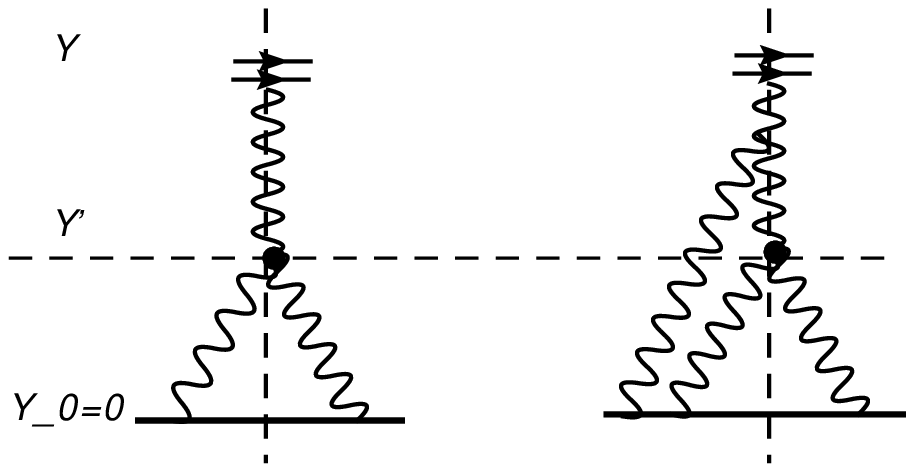,width=68mm,height=30mm}}
\end{minipage}
\caption{The examples of Pomeron diagrams that contribute to the process of single diffraction in the mean field approximation. The Pomeron, crossed by the dotted line,is the cut Pomeron.}
\label{sdxs} }


It is useful for our further discussions
to introduce a generating functional for multiparticle production with an arbitrary initial condition
\begin{eqnarray}\label{genfuncross}
M(Y-y;\{\gamma\},\{\gamma_{in}\})=1-Z(Y-y;\{u\},\{v\})|_{u=1-\gamma,v=1-\gamma_{in}}
\end{eqnarray}
for which boundary condition  \eq{MP1}  takes form of 
\begin{eqnarray}\label{bcM1}
M(Y-y;\{\gamma\},\{\gamma_{in}\})|_{\gamma_{in}=2\gamma}=2N(Y-y;\{\gamma\})
\end{eqnarray}

The generating functionals \eq{Zmulti} and \eq{genfuncross} should be supplemented by some initial condition
for calculating physical observables. The initial condition depends on the process we want to calculate. As it was already  mentioned, the linear equation for the generating functional will not depend on the initial condition.

As a simple example we calculate the cross section of single diffraction the initial condition for the generating functional has the following form
\beq \label{MPIC}
Z\Lb  Y;\{u\},\{v\}\Rb\,\,\,=\,\,v(r)
\eeq which means that we consider   diagrams which start with one cut Pomeron (see \fig{sdxs}).

The cross section of the single diffractive production with mass smaller than $\ln(M^2/m^2) \,\leq\,Y - Y'$ where $m$ is the nucleon mass, is equal to (see more details in Ref. \cite{LEPR0})
\beq \label{MPSD}
\sigma_{sd}\Lb Y, Y'\Rb\,\,=\,\, 1\,\,\,-\,\,Z\Lb Y - Y';\{u\};\{v\}  \Rb|^{v(r)}_{u(r)= 1 - N ( Y';r,b)\;,\;
v(r)= 1 - N^2 ( Y';r,b)}
\eeq
where $N(Y';r,b)$ reflects the fact that each Pomeron that cross line $Y'$ can develop its own tree `fan' of Pomerons. 

The substitution $v(r)= 1 - N^2 ( Y';r,b)$ is explained as follows. Any dipole that corresponds to $v(r)$ survives until $\tau=\infty$ ( is present on the unitarity cut) and thus can scatter both elastically and inelastically by $2N(Y';r,b)=|N(Y';r,b)|^2+G_{in}$ (see normalization of $\gamma_{in}=2\gamma$). However, we are interested in the single diffractive process with no particle production (inelastic scattering) below some rapidity $Y'$. That is the reason why 
we retain only the elastic part of the total cross section of a dipole present on the cut, namely, $N(Y';r,b)^2$.

\subsection{AGK cutting rules and vertices for dipole decays}

We wish to write the equation of \eq{Z2}-type for $P^m_n$, but first we need to determine the vertices of  decays of one $v$-dipole to two $u$-dipoles; one $v$-dipole to one $v$-dipole and one $u$-dipole; and one $v$-dipole to two $v$-dipoles.

We will use the AGK cutting rules
 of \eq{DC}-\eq{DDC} to find out  these vertices. According section 2.1 we expect the following rules 
for the different transition
\bea
\cpom \to \cpom\,+\,\cpom &\,\,\,\,\,\sim\,\,\,\,\,& \gamma^2_{in}\,\,\frac{\delta M}{\delta\,\gamma_{in}} \;\;
\label{vv}\\
\cpom \to \cpom\,+\,\pom &\,\,\,\,\,\sim\,\,\,\,\,& \gamma_{in}\,\gamma\,\,\frac{\delta M}{\delta\,\gamma_{in}} \;\;
\label{vu}\\
 \cpom \to \pom\,+\,\pom &\,\,\,\,\,\sim\,\,\,\,\,& \gamma\,\gamma\,\,\frac{\delta M}{\delta\,\gamma_{in}} \;\;
\label{uu}\\
\cpom \to \cpom  &\,\,\,\,\,\sim\,\,\,\,\,&  \gamma_{in} \,\,\frac{\delta M}{\delta\,\gamma_{in}}\;\;
\label{u}
 \eea
where $\cpom$ denotes the cut BFKL Pomeron.
At first sight, comparing Eqs.~(\ref{vv})-(\ref{u}) with the AGK cutting rules of Eqs.~(\ref{DC})-(\ref{DDC}) for  $\cpom \to \cpom\,+\,\cpom $ and $\cpom \to \cpom\,$ we have
 \bea
\cpom \to \cpom\,+\,\cpom &\sim & \frac{\bas}{2 \pi}\,\int\,d^2 r_2 \,K\Lb r_{10}| r_{12}, r_{02}\Rb 
\left\{\h \Lb\gamma_{in}(r_{12}) \,+\,\gamma_{in}(r_{02})\Rb^2 - \h\gamma^2_{in}(r_{10})\right\}\,\,\frac{\delta M(y;\{\gamma\},\{\gamma_{in}\})}{\delta\,\gamma_{in}}
\label{vv1}\\
\cpom \to \cpom  &\,\,\,\,\,\sim\,\,\,\,\,& \frac{\bas}{2 \pi}\,\int\,d^2 r_2 \,K\Lb r_{10}| r_{12}, r_{02}\Rb
\left\{ \gamma_{in}(r_{12}) + \gamma_{in}(r_{02}) - \gamma_{in}(r_{10})\right\} \,\,\frac{\delta M(y;\{\gamma\},\{\gamma_{in}\})}{\delta\,\gamma_{in}}
\label{u1}
 \eea
However, the AGK rules developed in \cite{LEPR0} were written for amplitudes, which are the solutions to the linear (BFKL and a generalized form of BFKL) equations. 
In the present study we are interested in the non-linear evolution and thus all quadratic terms of the same argument ($\gamma^2_{in}(12)$ etc.) are by definition absorbed in the corresponding linear terms, in other words
for the generating functional $M(y;\{\gamma\},\{\gamma_{in}\})$ functions $\gamma(r)$ and $\gamma_{in}(r)$ are arbitrary
and any quadratic term of the same argument can be absorbed in the definition of the corresponding linear term.  
 This means that the proper way to account for the 
 the AGK cutting rules is as follows
 \bea
\cpom \to \cpom\,+\,\cpom &\sim & \frac{\bas}{2 \pi}\,\int\,d^2 r_2 \,K\Lb r_{10}| r_{12}, r_{02}\Rb 
\gamma_{in}(r_{12}) \,\gamma_{in}(r_{02})\,\,\frac{\delta M(y;\{\gamma\},\{\gamma_{in}\})}{\delta\,\gamma_{in}(r_{10})}
\label{vv2}\\
\cpom \to \cpom\,+\,\pom &\sim &  -\,2\,\frac{\bas}{2 \pi}\,\int\,d^2 r_2 \,K\Lb r_{10}| r_{12}, r_{02}\Rb \left\{
\gamma_{in}(r_{12})\,\gamma(r_{12}) \,+\,\gamma_{in}(r_{02})\,\gamma(r_{02})\right\}\,\,\frac{\delta M(y;\{\gamma\},\{\gamma_{in}\})}{\delta\,\gamma_{in}(r_{10})}
\label{vu2}\\
 \cpom \to \pom\,+\,\pom &\sim &  \frac{\bas}{2 \pi}\,\int\,d^2 r_2 \,K\Lb r_{10}| r_{12}, r_{02}\Rb 
\gamma(r_{12})\,\gamma(r_{02})\,\,\frac{\delta M(y;\{\gamma\},\{\gamma_{in}\})}{\delta\,\gamma_{in}(r_{10})}
\label{uu2}\\
\cpom \to \cpom  &\,\,\,\,\,\sim\,\,\,\,\,& \frac{\bas}{2 \pi}\,\int\,d^2 r_2 \,K\Lb r_{10}| r_{12}, r_{02}\Rb 
\left\{ \gamma_{in}(r_{12}) + \gamma_{in}(r_{02}) - \gamma_{in}(r_{10})\right\} \,\,\frac{\delta M(y;\{\gamma\},\{\gamma_{in}\})}{\delta\,\gamma_{in}(r_{10})}
\label{u2}\\
\pom \to \pom\,+\,\pom &\sim &  \frac{\bas}{2 \pi}\,\int\,d^2 r_2 \,K\Lb r_{10}| r_{12}, r_{02}\Rb
\,\,\gamma(r_{12})\,\gamma(r_{02})\,\,\,\frac{\delta M(y;\{\gamma\},\{\gamma_{in}\})}{\delta\,\gamma(r_{10})}
\label{uuu} \\
\pom \to \pom &\sim &  \frac{\bas}{2 \pi}\,\int\,d^2 r_2 \,K\Lb r_{10}| r_{12}, r_{02}\Rb
\left\{ \gamma(r_{12}) + \gamma(r_{02}) - \gamma(r_{10})\right\}\frac{\delta M(y;\{\gamma\},\{\gamma_{in}\})}{\delta\,\gamma(r_{10})}
\label{u2u}
 \eea

From these equations we can easily to build the Pomeron Calculus for cut and uncut Pomerons
using the simple values: $\delta/\delta \gamma(r) $ and $\delta/\delta \gamma_{in}(r)$
are the annihilation operator for  uncut and cut Pomerons while the multiplication by $\gamma$ and $\gamma_{in}$ leads to a creation of uncut and cut Pomerons. It should be stressed that \eq{vv2}-\eq{u2u}
give a direct generalization of the equations for total cross section. In the latter we take into account the sum of all cuts.  This sum results in the equation in which transitions of \eq{uuu} and \eq{u2u} remain, but with opposite signs. It is clear that summing all the cuts  we obtain \eq{N3}. Therefore, we check a selfconsistence of our approach this way at the end of the present analysis.

\subsection{Linear functional equation}\label{sec:linfun}
\eq{vv2}-\eq{u2u} allow us to write the linear evolution equation for $N$ of \eq{MP}. It has the following form
\small
\bea \label{MPLEQ}
\small
&&\frac{ \partial M\Lb y, \{ \gamma\}, \{\gamma_{in} \}\Rb}{\partial\, y}\,\,= \\
&&\frac{\bas}{2 \pi}\,\int\,d^2 r_2 \,K\Lb r_{10}| r_{12}, r_{02}\Rb\left\{
\Lb \gamma(r_{12}) \,+\, \gamma(r_{02})\,-\,\gamma(r_{10}) \,-\,\gamma(r_{12})\,\gamma(r_{12})\Rb  \,
\frac{\delta M\Lb y, \{ \gamma\}, \{\gamma_{in} \}\Rb}{\delta\,\gamma(r_{10})} \,+\,\right.\notag \\
&&\left. +\,\, \Lb \gamma_{in}(r_{12}) \,+\, \gamma_{in}(r_{02})\,-\,\gamma_{in}(r_{10}) \,+\,\gamma_{in}(r_{12})\,\gamma_{in}(r_{02})\,-\,2\,\gamma_{in}(r_{12})\,\gamma(r_{02}) - 2\,\gamma_{in}(r_{02})\,\gamma(r_{12}) + 2\,\,\gamma(r_{12})\,\gamma(r_{02})\Rb\,\frac{\delta M\Lb y, \{ \gamma\}, \{\gamma_{in} \}\Rb}{\delta\,\gamma_{in}(r_{10})} \right\}\notag
\eea
\normalsize
The fact that we have different signs in front of terms $\gamma^2_{in}$, $ \gamma_{in} \gamma$ and $\gamma^2$ (see for example \eq{vu2}), makes the probabilistic interpretation very questionable.
However, it turns out that we can reduce \eq{MPLEQ} to a very simple equation with a very transparent probabilistic interpretation if we go back to the generating functional $Z \left(y;\{u\} ,\{v\}\right)$ where $u = 1 - \gamma$ and $v = 1 - \gamma_{in}$. As it was shown in \cite{LEPR0},  $Z \left(y;\{u\} ,\{v\}\right)$ can be written as a  functional of two other functions such that
 $Z \left(y;\{u \} ,\{v \}\right) \,\,=\,\,\tilde{Z} \left(y;\{u\} ,\{\xi\}\right)$, where we defined a new function $\xi(r)=2u(r)-v(r)$. 
 In terms of functions $u$ and $\xi$ the linear equation \eq{MPLEQ} has a simple form 
\bea \label{MPZLEQ}
&&\frac{\partial \tilde{Z} \left(y;\{u\} ,\{\xi \}\right)}{\partial y}\,\,=\,\,\frac{\bas}{2 \pi}\,\int\,d^2 r_2 \,K\Lb r_{10}| r_{12}, r_{02}\Rb \times \\
&&\left\{ \Lb u(r_{12})\,u(r_{02})\,-\,u(r_{10}) \Rb \frac{\delta  \tilde{Z} \left(y;\{u\} ,\{\xi\}\right)}{\delta  u(r_{10})}\,\,+\,\,\Lb \xi(r_{12})\,\xi(r_{02})\,-\,\xi(r_{10}) \Rb\,\frac{\delta  \tilde{Z} \left(y;\{u\} ,\{\xi \}\right)}{\delta  \xi(r_{10})}\right\}
\notag
\eea
\eq{MPZLEQ} has  a transparent probabilistic interpretation, since the second term at the r.h.s. of the equation is of  the same structure as the first one and describes the Markov chain for the decay of the dipole which corresponds to $ 2u-v$.  This physical meaning becomes clear if we recall that
 in the generating functional approach the functions $u$ or $v$ correspond to the creation operator for dipole that does not  or does survive till time $\tau=\infty$, respectively (for corresponding annihilation operators  we have $\delta/\delta u$ and $\delta/\delta v$).

The  solution to \eq{MPZLEQ} with the initial condition of \eq{MPIC} is found as 
\beq \label{MPZFIN11}
\tilde{Z} \left(y; \{u\} ,\{\xi \}\right)|^{2u(r)-\xi(r)}\,\,\,=\,\,\,2\,Z_0\Lb y; \{u\}\Rb|^{u(r)} \,\,-\,\,
Z_0\Lb y; \{\xi\}\Rb
|^{\xi(r)}\eeq

One can easily check that the solution \eq{MPZFIN11} satisfies boundary conditions \eq{MPZBC} and  \eq{MP1}.
 It is very instructive to compare this with the explicit form of the solution in the toy model found in
Ref.~\cite{LEPR0}.

\section{Non-linear evolution equation}\label{sec:nonlin}
In this section we rewrite the linear functional first order differential equation \eq{MPZLEQ} as the non-linear equation for the scattering amplitude using the initial condition for the generating functionals $Z_0$ and $Z$, given by \eq{ZIC} and \eq{MPIC}, respectively.   
The easiest way is to go back to \eq{MPLEQ} and using the initial condition for 
  $N(y;\{\gamma\})$ (see \eq{initN}) and $M(y;\{\gamma\},\{\gamma_{in}\})$, to  write the non-linear equation in analogy with transition from \eq{Z3} to \eq{Z4}. The initial condition for the cross section 
 $M\Lb y; \{ \gamma\}, \{\gamma_{in} \}\Rb$ is easily obtained from \eq{genfuncross} and \eq{MPIC} in the form of 
\beq \label{initM}
M(y=0;\{\gamma\},\{\gamma_{in}\})=\gamma_{in}(r)
\eeq
Plugging \eq{initN} and \eq{initM} into the linear equation \eq{MPLEQ} we get the non-linear equation   
\bea \label{MPNEQ11}
&& \frac{\partial M(y;\{\gamma\},\{\gamma_{in}\})|^{\gamma_{in}(r_{10})}}{\partial y}\,=\,
\,\,\frac{\bas}{2 \pi}\,\int\,d^2 r_2 \,K\Lb r_{10}| r_{12}, r_{02}\Rb \times \\
&&\left\{M(y;\{\gamma\},\{\gamma_{in}\})|^{\gamma_{in}(r_{12})}\,+\,\begin{scriptsize}\begin{footnotesize}\end{footnotesize}\end{scriptsize}
M(y;\{\gamma\},\{\gamma_{in}\})|^{\gamma_{in}(r_{20})}\,-
\,M(y;\{\gamma\},\{\gamma_{in}\})|^{\gamma_{in}(r_{10})}\,\right.\notag \\
&& \left.+
M(y;\{\gamma\},\{\gamma_{in}\})|^{\gamma_{in}(r_{12})}M(y;\{\gamma\},\{\gamma_{in}\})|^{\gamma_{in}(r_{20})}
\,-\,2\,M(y;\{\gamma\},\{\gamma_{in}\})|^{\gamma_{in}(r_{12})}N(y;\{\gamma\})|^{\gamma(r_{20})} \right. \notag \\
&&\left. -\,2\,
N(y;\{\gamma\} )|^{\gamma (r_{12})}
M(y;\{\gamma\},\{\gamma_{in}\})|^{\gamma_{in}(r_{20})}
\,+\, 2
N(y;\{\gamma\} )|^{\gamma (r_{12})}
N(y;\{\gamma\} )|^{\gamma (r_{20})}
\right\}\notag
 \eea
With the help of \eq{N11}, \eq{Nfundef} \eq{MP} \eq{genfuncross} we recast  \eq{MPNEQ11} 
in a familiar form  of
\bea \label{MPNEQ}
&& \frac{\partial M(y;r_{10},b)}{\partial Y}\,=\,
\,\,\frac{\bas}{2 \pi}\,\int\,d^2 r_2 \,K\Lb r_{10}| r_{12}, r_{02}\Rb \left\{M(y;r_{12},b)\,+\,\begin{scriptsize}\begin{footnotesize}\end{footnotesize}\end{scriptsize}
M(y;r_{20},b)\,-
\,M(y;r_{10},b)\,\right.  \\
&& \left.+
M(y;r_{12},b)M(y;r_{20},b)
\,-\,2\,M(y;r_{12},b)N(y;r_{20},b) -\,2\,
N(y;r_{12},b)
M(y;r_{20},b)
\,+\, 2
N(y;r_{12},b)
N(y;r_{20},b)
\right\}\notag
 \eea
\eq{MPNEQ} describes all multiparticle production processes and has the same form as the equation for the  cross section of diffractive production found in Ref. \cite{KLDD}.

 The difference between various processes manifests itself only in the  different initial conditions. For example for the diffractive production  the initial condition 
has the following form (see \fig{sdxs} and \eq{MPSD})
\beq \label{MPSD1}
\,M \left(Y=Y';r,b\right)=\,\,N^2_0 \left(Y' ,r,b\right)
\eeq

where $N_0$ is the amplitude of elastic scattering at rapidity $Y =Y'$. \eq{MPSD1} can be directly obtained from the corresponding condition for the generating functional \eq{MPSD} noting that $M \left(Y=Y';r,b\right)$ has meaning 
of the total cross section. This way we show that we can describe a process either in terms  of the linear equation for the generating functional or a  non-linear evolution equation for the scattering amplitude (total cross section).  It should be noted that the generating functional approach is more general and allows more freedom in 
the choice of the initial condition.

If we want to find a cross section with the $k$-recoiled nucleons in dipole nucleus interaction we need to calculate
\beq \label{XSK}
\sigma^{(k)}\Lb Y,r\Rb \,\,\,=\,\,\frac{1}{k!}\,\prod^k_{i = 1}\,\gamma_{in}(r_i)\left(\frac{\delta}{\delta \gamma_{in}(r_i)}\,M \left(Y;\{\gamma\} ,\{\gamma_{in} \}\right)|^{\gamma_{in}(r)}\right)|_{\gamma_{in}=0}
\eeq
where $\gamma(r)$ is the low energy elastic amplitude and $\gamma_{in}(r) = 2 \gamma(r)$ at low energy ($Y_0$).

This cross section can be found as the solution to non-linear equation with the initial condition
\beq \label{INCFM}
\sigma^{(k)}\Lb Y_0,r;b\Rb\,\,=\,\,\frac{1}{k!}\,e^{ - \Omega\Lb Y_0 ,r;b\Rb}
\,\Omega\Lb Y_0 ,r;b\Rb
\eeq
where $\Omega\Lb Y_0 ,r;b\Rb = \sigma_{\mbox{dipole-proton}}(Y=Y_0,r)\,T_A\Lb b\Rb$ 
with $\sigma_{\mbox{dipole-proton}}(Y=Y_0,r)$ is the cross section of  dipole - nucleon interaction at low energy  being equal to  $\sigma_{\mbox{dipole-proton}}(Y=Y_0,r) = 
2\int d^2 b' \, N_0(Y_0,r,b')$ . The function $T_A\Lb b\Rb$ is the optical width of nucleus which gives the  number of nucleons at given value of impact parameter $b$.  We assume that $Y_0$ is large enough to use the Glauber approach. The derivation of \eq{INCFM}  is given, for example,  in Ref. \cite{Kormilitzin:2007je}. 

\section{Conclusions}
The main result of the paper is  two equations for the generating functional for multiparticle production: 
the linear equation \eq{MPLEQ} and \eq{MPZLEQ}, and the non-linear equation  \eq{MPNEQ}.
The linear equations have an advantage of being correct for any initial condition, while the non-linear equation describes the process which starts from the exchange of a single Pomeron at low energies.

The non-linear equation has the same form as the equation for diffractive production that has been proved in Ref. \cite{KLDD} and confirms in Refs. \cite{KLW,HIMST,HWS}. We hope that here we give a more transparent and physically motivated derivation for diffractive production and generalize the approach to other processes of multiparticle generation. It should be stressed that the processes that we considered here are totally inclusive in the sense that we do not measure a particular particle in these processes. For example, our equations cannot describe the multiparticle inclusive correlations since we do not have AGK cutting rules for vertices with the emission of gluon (see \fig{agkver}).

In this paper we consider only the multiparticle processes in MFA, but we hope to use Mueller-Patel-Salam-Iancu approach \cite{MPSI} to calculate these processes taking into account the Pomeron loops in spirit of
  the approach suggested in Refs. \cite{LEPR0,LMP}.

\section* {Acknowledgements}
We are grateful to Joachim Bartels, Errol Gotsman, Yuri Kovchegov, Lev Lipatov, Uri Maor and Kirill Tuchin  for fruitful  discussions on the subject. 
This research was supported  in part by the Israel Science Foundation, founded by the Israeli Academy of Science 
and Humanities, by BSF grant $\#$ 20004019 and by 
a grant from Israel Ministry of Science, Culture and Sport and 
the Foundation for Basic Research of the Russian Federation.  

~


\begin{thebibliography}{99}

\bibitem{GLR}
L. V. Gribov, E. M. Levin and M. G. Ryskin, {\it Phys. Rep.}\,
{\bf 100}, 1 (1983).

\bibitem{MUQI}
A. H. Mueller and J. Qiu,  {\it Nucl. Phys.} \,{\bf B 268}\,427
(1986) .
\bibitem{MV}
L. McLerran and R. Venugopalan, {\it  Phys. Rev.}  {\bf D 49},2233,
3352  (1994); {\bf D 50},2225 (1994); {\bf D 53},458 (1996); {\bf
D 59},09400
(1999).

\bibitem{B}
I.~Balitsky,
[arXiv:hep-ph/9509348];\,\,
{\it Phys.\ Rev.} {\bf D60}, 014020 (1999)
[arXiv:hep-ph/9812311]\,\,\,\,


\bibitem{K}
Y.~V.~Kovchegov,
{\it Phys.\ Rev.}  {\bf D60}, 034008  (1999),
[arXiv:hep-ph/9901281].

\bibitem{KLDD}
  Y.~V.~Kovchegov and E.~Levin,
  Nucl.\ Phys.\  B {\bf 577} (2000) 221
  [arXiv:hep-ph/9911523].
\bibitem{AGK}
  V.~A.~Abramovsky, V.~N.~Gribov and O.~V.~Kancheli,
  Yad.\ Fiz.\  {\bf 18}, 595 (1973)
  [Sov.\ J.\ Nucl.\ Phys.\  {\bf 18}, 308 (1974)].


\bibitem{MUCD}
A.~H.~Mueller, {\it Nucl.\ Phys.} {\bf B415}, 373 (1994);
{\it ibid}  {\bf B437}, 107 (1995).


  \bibitem{AGKQCD}
    M.~Salvadore, J.~Bartels and G.~P.~Vacca,
  {\it ``Multiple interactions and AGK rules in pQCD,''}
  arXiv:0709.3062 [hep-ph];\,\,\,
  F.~Gelis and R.~Venugopalan,
  Nucl.\ Phys.\  A {\bf 782} (2007) 297,
   {\bf 785} (2007) 146],
  [arXiv:hep-ph/0608117];\,\,\,
   J.~Bartels, M.~Salvadore and G.~P.~Vacca,
  Eur.\ Phys.\ J.\  C {\bf 42} (2005) 53
  [arXiv:hep-ph/0503049];\,\,\,
  J.~Bartels and M.~G.~Ryskin,
  Z.\ Phys.\  C {\bf 76} (1997) 241
  [arXiv:hep-ph/9612226];\,\,\,
  D.~Treleani,
  Int.\ J.\ Mod.\ Phys.\  A {\bf 11} (1996) 613.


\bibitem{K1}
  Y.~V.~Kovchegov,
  Phys.\ Rev.\  D {\bf 64}, 114016 (2001)
  [Erratum-ibid.\  D {\bf 68}, 039901 (2003)]
  [arXiv:hep-ph/0107256].


\bibitem{K2}
  Y.~V.~Kovchegov and K.~Tuchin,
  Phys.\ Rev.\  D {\bf 65}, 074026 (2002)
  [arXiv:hep-ph/0111362].



\bibitem{K3}
  J.~Jalilian-Marian and Y.~V.~Kovchegov,
  Phys.\ Rev.\  D {\bf 70}, 114017 (2004)
  [Erratum-ibid.\  D {\bf 71}, 079901 (2005)]
  [arXiv:hep-ph/0405266].
\bibitem{BFKL}
 E. A. Kuraev, L. N. Lipatov, and F. S. Fadin, {\it  Sov. Phys.
JETP}
                {\bf 45}, 199 (1977); \,\,\,
Ya. Ya. Balitsky and L. N. Lipatov,
               {\it   Sov. J. Nucl. Phys.}\, {\bf 28}, 22 (1978).
\bibitem{LEPR}
E.~Levin and A.~Prygarin,
{\it `` Inclusive gluon production in the dipole approach: AGK cutting rules"}.




\bibitem{LEPR0}
 E.~Levin and A.~Prygarin,
  Eur.\ Phys.\ J.\  C {\bf 53} (2008) 385
  [arXiv:hep-ph/0701178].
\bibitem{LL}
E.~Levin and M.~Lublinsky,
  Nucl.\ Phys.\  A {\bf 763} (2005) 172
  [arXiv:hep-ph/0501173];\,\,\,
  Phys.\ Lett.\  B {\bf 607} (2005) 131
  [arXiv:hep-ph/0411121];\,\,\,
  Nucl.\ Phys.\  A {\bf 730} (2004) 191
  [arXiv:hep-ph/0308279].


\bibitem{KLW}
 A.~Kovner, M.~Lublinsky and H.~Weigert,
  Phys.\ Rev.\  D {\bf 74}, 114023 (2006)
  [arXiv:hep-ph/0608258].


\bibitem{HIMST}
Y.~Hatta, E.~Iancu, C.~Marquet, G.~Soyez and D.~N.~Triantafyllopoulos,
  Nucl.\ Phys.\  A {\bf 773} (2006) 95
  [arXiv:hep-ph/0601150].
\bibitem{HWS}
 M.~Hentschinski, H.~Weigert and A.~Schafer,
  Phys.\ Rev.\  D {\bf 73} (2006) 051501
  [arXiv:hep-ph/0509272].
\bibitem{MPSI}
A.~H.~Mueller and B.~Patel,
{\it Nucl.\ Phys.}  {\bf B425}, 471 (1994);\,\,\,
A.~H.~Mueller and G.~P.~Salam,
{\it  Nucl.\ Phys.}\,  {\bf B475}, 293 (1996),
[arXiv:hep-ph/9605302];\,\,\,\,G.~P.~Salam,
{\it   Nucl.\ Phys.}\,  {\bf B461}, 512 (1996);\,\,\,
E.~Iancu and A.~H.~Mueller,
{\it  Nucl.\ Phys.}\,  {\bf A730} (2004) 460, 494,
[arXiv:hep-ph/0308315],[arXiv:hep-ph/0309276].
\bibitem{LMP}
  E.~Levin, J.~Miller and A.~Prygarin,
  {\it ``Summing Pomeron loops in the dipole approach,''} Nucl.Phys. A (in press),
  arXiv:0706.2944 [hep-ph].

\bibitem{Kormilitzin:2007je}
  A.~Kormilitzin,
  arXiv:0707.2202 [hep-ph].


\end{thebibliography}
\end{document}